\documentclass[]{spie}  

\usepackage{amsmath,amsfonts,amssymb}
\usepackage{graphicx}
\usepackage{multirow}
\usepackage{wrapfig}
\usepackage{subcaption}
\usepackage{sidecap}
\usepackage{siunitx}
\usepackage{tabularx}
\usepackage{array}
\usepackage[table, x11names]{xcolor}
\usepackage[colorlinks=true, allcolors=blue]{hyperref}

\title{Liger at W.M. Keck Observatory: imager structural analysis, fabrication, and characterization plan}

\author[a]{James Wiley}
\author[b]{Aaron Brown}
\author[c]{Renate Kupke}
\author[d]{Maren Cosens}
\author[a,b]{Shelley A. Wright}
\author[b]{Jérôme Maire}
\author[e]{Kenneth Magnone}
\author[e]{Evan Kress}
\author[e]{Eric Wang}
\author[e]{Chris Johnson}
\author[e]{James E. Larkin}
\author[e]{Michael P. Fitzgerald}
\author[f]{Marc Kassis}
\author[g]{Tucker Jones}

\affil[a]{Department of Physics, University of California San Diego, USA}
\affil[b]{Department of Astronomy \& Astrophysics, University of California San Diego, USA}
\affil[c]{Department of Astronomy \& Astrophysics, University of California Santa Cruz, USA}
\affil[d]{Carnegie Science, The Observatories, Pasadena, CA, USA}
\affil[e]{Department of Physics \& Astronomy, University of California Los Angeles, USA}
\affil[f]{W.M. Keck Observatory, Waimea, HI, USA}
\affil[g]{Department of Physics, University of California Davis, USA}

\authorinfo{Send correspondence to jhwiley@ucsd.edu}

\begin{document} 
\maketitle

\begin{abstract}
Liger is an adaptive optics (AO) fed imager and integral field spectrograph (IFS) designed to take advantage of the Keck All-sky Precision Adaptive-optics (KAPA) upgrade to the Keck I telescope. Liger adapts the design of the InfraRed Imaging Spectrograph (IRIS) for the Thirty Meter Telescope (TMT) to Keck by implementing a new imager and re-imaging optics. The performance of the imager is critical as it sequentially feeds the spectrograph and contains essential components such as the pupil wheel, filter wheel, and pupil viewing camera. We present the design and structural analysis of the Liger imager optical assembly including static, modal, and thermal simulations. We present the fabrication as well as the full assembly and characterization plan. The imager will be assembled bench-top in a clean room utilizing a coordinate-measuring machine (CMM) for warm alignment. To ensure optimal performance, the imager will be characterized in a test cryostat before integration with the full Liger instrument. This comprehensive approach to characterization ensures the precision and reliability of the imager, enhancing the observational capabilities of Liger and W.M. Keck Observatory.
\end{abstract}


\section{Liger Imager Overview}\label{sec:1}

Liger is a second generation, adaptive optics (AO) fed imager and integral field spectrograph (IFS) designed to take advantage of the Keck All-Sky Precision Adaptive-optics (KAPA)\cite{Wizinowich2022} upgrade to the Keck I AO system\cite{Wright2019}. Liger builds upon the success and foundation of the OH-Suppressing Infra-Red Imaging Spectrograph
(OSIRIS)\cite{Larkin2006} which was critical to many science campaigns including the Nobel prize winning research of the discovery of the supermassive black hole at the center of the Milky Way\cite{Ghez2008}. Liger also serves as a pathfinder instrument for the Thirty Meter Telescope (TMT) first light instrument, the InfraRed Imaging Spectrograph (IRIS)\cite{Larkin2020}, as Liger adapts the same spectrograph design to Keck by implementing a new imager and re-imaging optics (RIO)\cite{Wiley2022}\cite{Cosens2022}. Liger is unique in that it provides higher spectral resolving power (R=4,000-10,000), wider wavelength coverage (0.81-2.45 \textmu m), and larger fields of view (up to 13.2"$\times$6.8") than any current AO-fed IFS.

\begin{figure}[t!]
\begin{center}
  \includegraphics[width=1\linewidth,trim={0.0in 0.0in 0.0in 0.0in}]{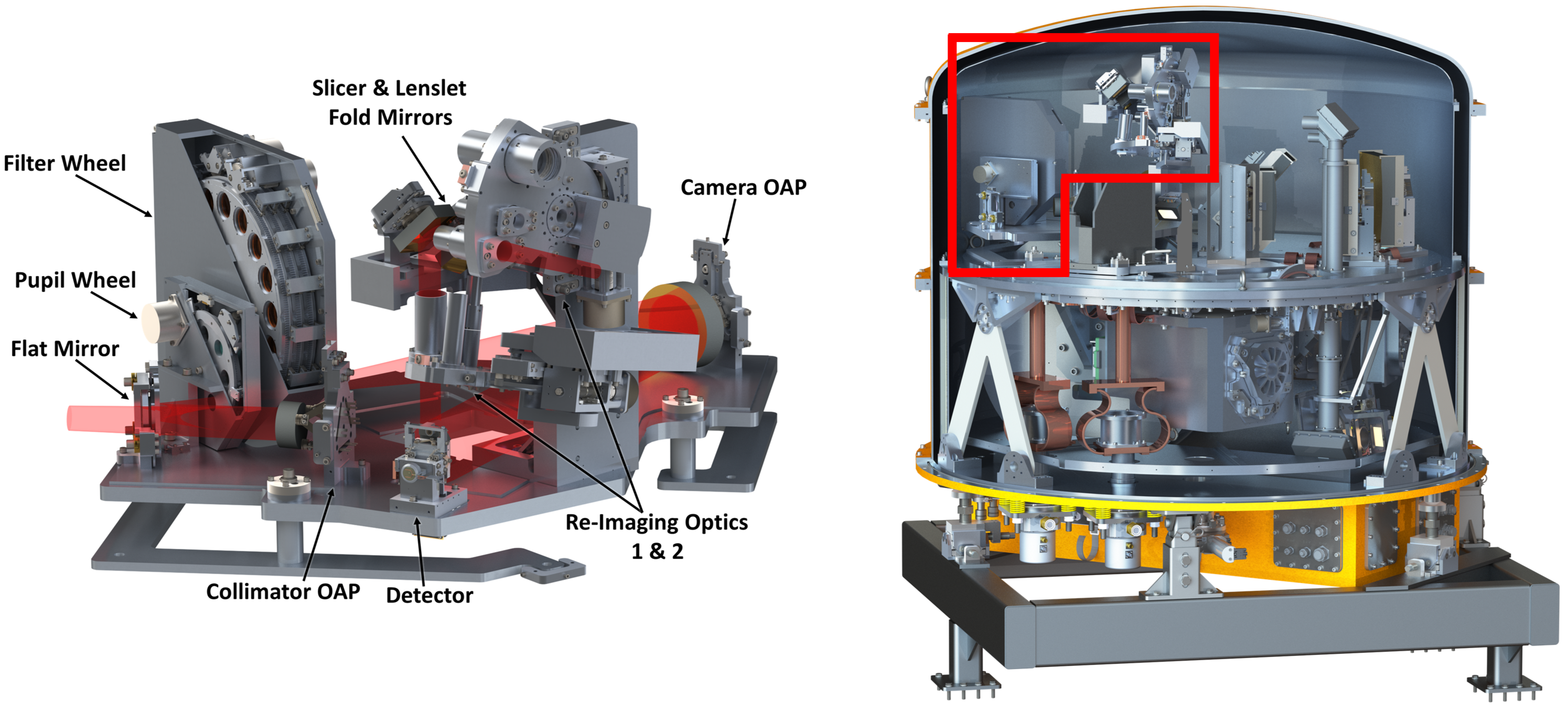}
  \end{center}
\caption[Figure1]{Left: Rendering of the Liger imager optical assembly showing every major component except the pupil viewing camera which is hidden by the RIO tower. This view has baffling removed and cuts in the filter wheel and pupil wheel housings. Right: Rendering of the inside of the Liger science cryostat highlighting the location of the imager. This view includes cuts through the vacuum and cold shield and also has baffling removed. In both renderings, the appearance of each component represents its material as the actual appearance will be mostly black to suppress internal reflections.}
\label{fig:1}
\end{figure}

The Liger instrument consists of multiple optical subsystems inside a cryogenic dewar mounted behind the AO bench on the Keck I Nasmyth platform. The first subsystem encountered by the AO beam as it enters the instrument is the imager which transfers the beam to the imaging detector and pick-off mirrors that send the rest of the beam to both spectrograph subsystems. Figure \ref{fig:1} shows the inside of the Liger cryostat highlighting the imager subsystem. The full instrument will be the largest cryogenic dewar at Keck at around 1.8 meters tall and 1.4 meters in diameter and weighing close to 6,000 lbs. The imager is a small portion of the overall instrument but it alone requires a large vacuum chamber and robust in-lab setup for characterization\cite{Wiley2020}.

The imager is optically simple, consisting of two off axis parabolic (OAP) mirrors and a fold mirror, but the precision and alignment are critical to meet Liger's science goals of high precision photometry, astrometry, and Point Spread Function (PSF) reconstruction\cite{Kupke_2024}. The imager maintains an optimal PSF by suppressing diffraction spikes and radiation with a rotating pupil wheel that can serve as both a Lyot stop and cold stop\cite{Cosens2020}. The pupil wheel is aligned with a pupil viewing camera that is rotated out of the beam path during normal operation and
\begin{table}[b!]
\centering
\caption{Optical design requirements derived from science requirements for the imaging camera.}
\label{tab:1}
\begin{tabularx}{\textwidth}{|>{\centering\arraybackslash}m{0.2\textwidth}|>{\centering\arraybackslash}m{0.2\textwidth}|>{\centering\arraybackslash}m{0.523\textwidth}|}
  \hline
  \textbf{Requirement} & \textbf{Value} & \textbf{Example Limiting Science Cases} \\
  \hline
  \multirow{2}{*}{Wavelength Range} & \multirow{2}{*}{0.81 - 2.45 \textmu m} & \multirow{2}{0.512\textwidth}{\centering Simultaneous PSF characterization for spectroscopic cases including stellar astrophysics and lensed QSOs} \\
  & & \\
  \hline
  \multirow{2}{*}{Field of View} & \multirow{2}{*}{20.5" $\times$ 20.5"} & \multirow{2}{0.512\textwidth}{\centering Proper motions of extended systems, astrometric and PSF references; mapping solar system objects} \\
  & & \\
  \hline
  \multirow{2}{*}{Spatial Sampling} & \multirow{2}{*}{10 mas per pixel} & \multirow{2}{0.512\textwidth}{\centering Nyquist sampling of PSF for Galactic Center; cosmology; stellar proper motions; and exoplanets} \\
  & & \\
  \hline
  \multirow{2}{*}{Image Quality} & \multirow{2}{*}{$< 50$ nm RMS WFE} & \multirow{2}{0.512\textwidth}{\centering Diffraction-limited imaging at shortest possible wavelengths for lensed QSOs and Galactic Center} \\
  & & \\
  \hline
  \multirow{2}{*}{Efficiency} & \multirow{2}{*}{$>40$\%} & \multirow{2}{0.512\textwidth}{\centering Exoplanet direct imaging with precision astrometry and proper motions} \\
  & & \\
  \hline
 \end{tabularx}
\end{table}
focuses the re-imaged pupil plane on the imager detector during alignment\cite{Wiley2022}. The imager H2RG detector is held on the same mount that holds the IFS pick-off mirrors and allows for individual alignment of each\cite{Cosens2022}. The imager provides a low wavefront error (WFE) of 50 nm over the whole detector and pick-off mirror faces with a clean and static optical distortion solution. The optical design requirements derived from the science requirements for the Liger imager are given in table \ref{tab:1}.

This paper describes the structural analysis of the final design of the imager optical assembly, showing that it meets the requirements for Keck observatory, and details the fabrication and in lab characterization setup. For the overall design and fabrication status of the Liger instrument, see (\citenum{Wright_2024}). For the optical design and alignment strategy of the Liger IFS, see (\citenum{Kupke_2024}). For the assembly, integration, and testing, see (\citenum{Brown_2024}). This paper is organized as follows: \S\ref{sec:1} gives an overview of the Liger instrument and details the importance of the precision and alignment of the imager subsystem. \S\ref{sec:2} describes the structural analysis performed on the Liger imager optical assembly. This section includes modal, static, and thermal analyses which show that the imager as a whole meets requirements. \S\ref{sec:3} details the in lab experimental setup and characterization plan for the Liger imager, as well as the current fabrication status of the project. \S\ref{sec:4} summarizes the latest design and explains the necessary future steps for completion of the Liger imager.

\section{Imager Structural Analysis}\label{sec:2}


\begin{figure}[b]
\begin{center}
\minipage{0.499\textwidth}
\begin{center}
  \includegraphics[width=0.999\linewidth,trim={0.0in 0.0in 0.0in 0.0in}]{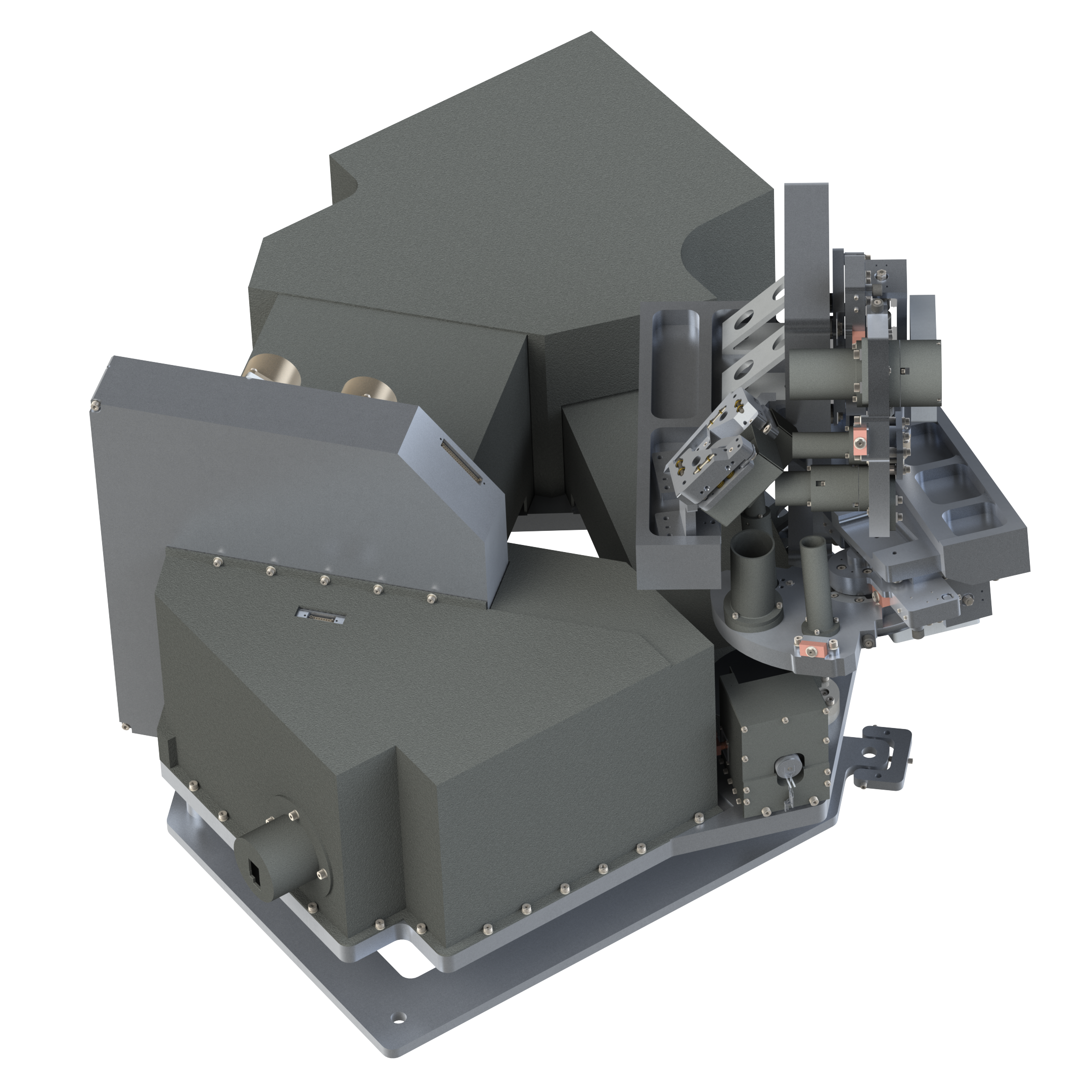}
  \end{center}
\endminipage\hfill
\minipage{0.499\textwidth}
\begin{center}
  \includegraphics[width=.999\linewidth,trim={0.0in 0.0in 0.0in 0.0in}]{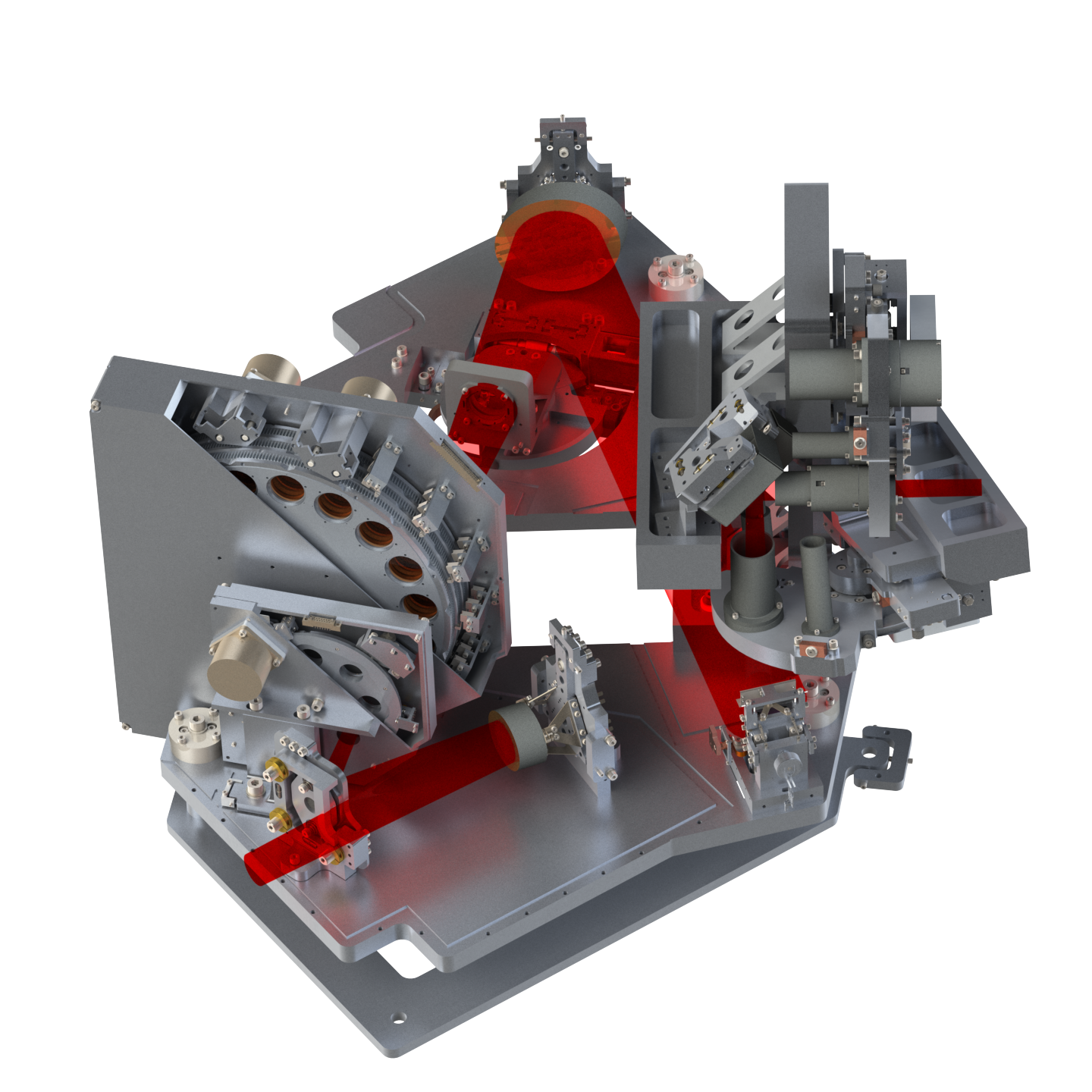}
\end{center}
\endminipage\hfill
\caption[Figure2]{Renderings of the Liger imager optical assembly showing the full assembly on the left and with baffling removed on the right. Following the beam as it enters the instrument from the bottom left, the components encountered in order are the collimator OAP, fold mirror, pupil wheel, filter wheel, pupil viewing camera, Camera OAP, the imager detector \& pick-off mirrors, and the RIO 1 \& 2 tower.}
\label{fig:2}
\end{center}
\end{figure}

The Liger imager provides a 20.5"$\times$20.5" field of view at 10 mas scales with sensitivity over the 0.81-2.45 \textmu m range. It provides high precision astrometric and photometric data to accompany IFS data for every observation. The imager data is also used along with KAPA AO telemetry to analytically reconstruct the average PSF for each observation, a process known as PSF-Reconstruction (PSFR)\cite{Sabhlok2024}. PSFR allows for image deconvolution, improving the spatial resolution and sensitivity of each observation. Reconstructed PSFs are also used for different science cases such as subtracting the central PSF when observing a Quasi-Stellar Object (QSO) to resolve its host galaxy\cite{Vayner2016}.

The imager is the first subsystem seen by the beam path as it exits the AO bench and enters the Liger instrument. It is critical to the overall Liger system as it transfers the pristine beam from the AO bench to both IFS subsystems.  The imager achieves this pristine transfer of the beam by utilizing two OAP mirrors on custom mounts for precision alignment. A fold mirror on a tip-tilt mount keeps the light path within the Liger footprint. The imager optical plate contains a rotating pupil wheel with 7 unique slots and a filter wheel that can hold up to 51 different filters\cite{Cosens2020}. A pupil viewing camera is located between the filter wheel and camera OAP and rotates in and out of the beam path to focus the image of the entrance pupil of the telescope on the imager detector for alignment of the pupil mask\cite{Wiley2022}. The imager detector is mounted on a stage that also holds the IFS pick-off mirrors and allows for independent alignment at the imager focal plane\cite{Cosens2022}. The RIO tower which holds the re-imaging doublets for the IFS is mounted to the imager optical plate and positions the doublets above the pick-off mirrors. Baffling is placed over the entire beam path, and the whole assembly is mounted on an adapter frame that is used for alignment within the full Liger instrument\cite{Wiley2022}. Figure \ref{fig:2} shows the imager assembly and all its components both with and without baffling.

\begin{table}[t]
\centering
\caption{Individual imager components and the Electronic Identifier (EID) for the publication describing their design and analysis.}
\label{tab:2}
\begin{tabular}{|c|c|} 
\hline
\textbf{Imager Component} & \textbf{SPIE EID}\\\hline
OAP 1, 2 \& Fold Mirror \cite{Wiley2020} & \href{https://doi.org/10.1117/12.2561837}{1144758}\\\hline
Pupil \& Filter Wheels \cite{Cosens2020} & \href{https://doi.org/10.1117/12.2561837}{114474X}\\\hline
Pupil Viewing Camera \cite{Wiley2022} & \href{https://doi.org/10.1117/12.2561837}{1218464}\\\hline
Detector Mount \cite{Cosens2022} & \href{https://doi.org/10.1117/12.2561837}{121845Y}\\\hline
Re-Imaging Optics 1 \& 2 \cite{Wiley2022} & \href{https://doi.org/10.1117/12.2561837}{1218464}\\\hline
Optics \& Adapter Plates \cite{Wiley2022} &  \href{https://doi.org/10.1117/12.2561837}{1218464}\\\hline
Baffling \cite{Wiley2022} & \href{https://doi.org/10.1117/12.2561837}{1218464}\\\hline
Test Chamber \cite{Wiley2020} & \href{https://doi.org/10.1117/12.2561837}{1144758}\\\hline
\end{tabular}
\end{table}

\begin{wrapfigure}{r}{0.68\textwidth}
  \begin{center}
    \includegraphics[width=0.68\textwidth]{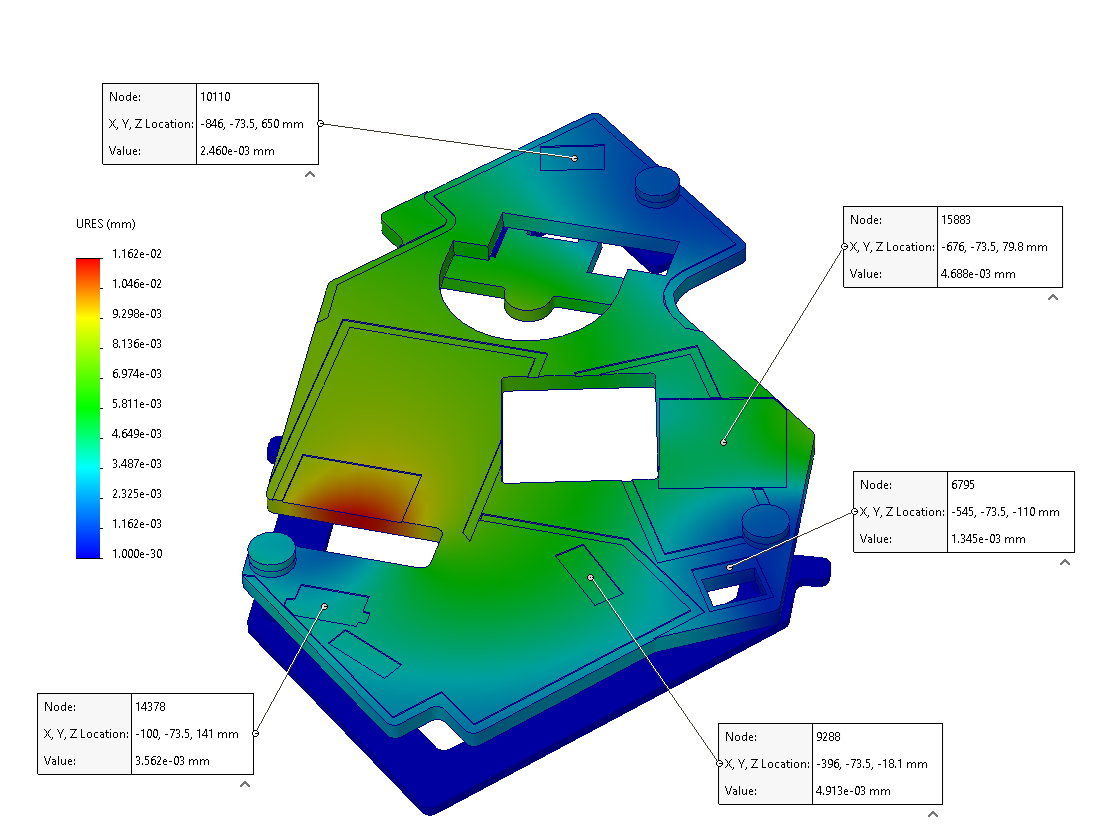}
  \end{center}
  \caption[Figure3]{SolidWorks static Simulation result showing the displacement of the imager optical assembly under gravity. The max displacement is about 12 \textmu m with probes showing the displacement at the OAP 1, pupil wheel, OAP 2, RIO tower, and detector mount locations.}
\label{fig:3}
\end{wrapfigure}

The entire imager model is analyzed in SolidWorks Simulation to ensure each component can be aligned within the required tolerance (table \ref{tab:3}), that the imager will operate outside the vibration keep out range of 8-80 Hz when installed at Keck, and that there are no large stresses or displacements from the thermal gradient when cooled. Each individual component has been analyzed to ensure proper performance in previous SPIE papers linked in table \ref{tab:2}. This section is split into three subsections for each analysis performed on the imager. \S\ref{sec:2.1} describes the static analysis of the imager showing it will survive stresses due to gravity and shipping and that displacements due to gravity are within the range of adjustability for each component. \S\ref{sec:2.2} describes the modal analysis of the imager and shows that it meets the requirements set by Keck. \S\ref{sec:2.3}  Shows the thermal analysis of the Liger imager and the expected temperature gradient across the optical plate.

\subsection{Static Analysis}\label{sec:2.1}

The full CAD model of the Liger imager optical assembly was loaded into SolidWorks Simulation for the static finite element analysis (FEA). This analysis is performed to ensure the structural integrity of the imager assembly and that no component mounting locations will deviate significantly from their expected position under gravity.

To simplify the simulations for optimal performance, the pupil and filter wheels were deconstructed by removing their interior components and simulating their weight as a point mass. Point masses were similarly placed at the center of gravity of OAP 1, OAP 2, the fold mirror, the pupil viewing camera, the RIO tower, and the imager detector. Complicated geometries were simplified for computational efficiency and improved convergence and stability. 

The connections within the simulation model were represented by bonding the opposing faces of fastened components. The base of the adapter frame was designated as the fixture point. Gravity was the sole external force applied for these simulations, and an average mesh density was used.

The results of the static analysis confirm the structural integrity of the assembly. The maximum stress in the assembly is 3 MPa located around the adapter pedestals between the optical and adapter plates. This stress is about a factor of 80 below yield for 6061 T6 aluminum which shows the robustness of the setup and that the assembly meets the shipping requirements to survive vertical shocks. 

\begin{figure}[b]
\begin{center}
  \includegraphics[width=1\linewidth,trim={0.0in 0.0in 0.0in 0.0in}]{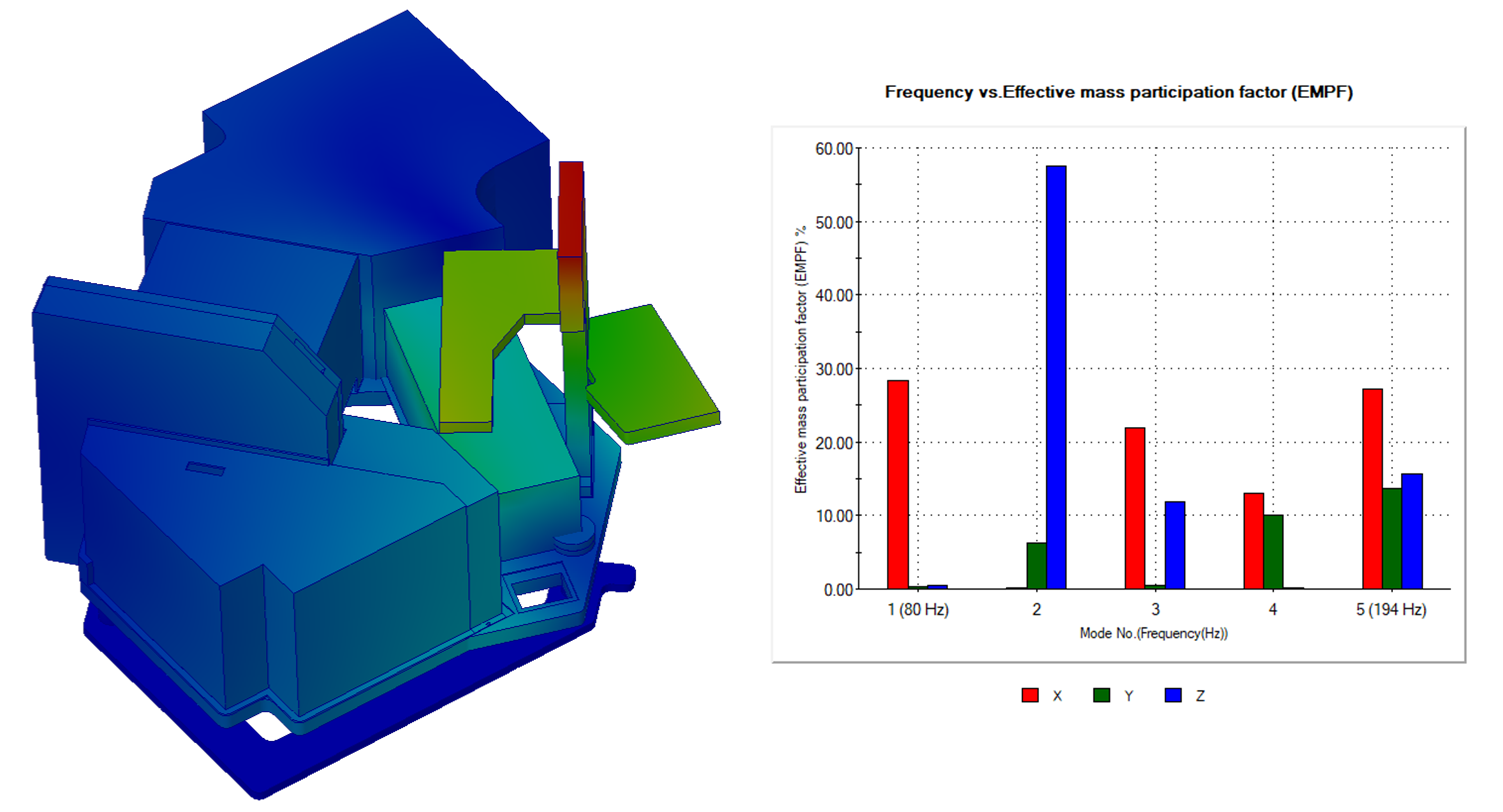}
  \end{center}
\caption[Figure4]{Solidworks Simulation result of the natural frequencies of the imager optical assembly. The view on the left is of the first resonance at 80 Hz which is on the edge of the Keck keep-out range of 8-80 Hz. The graph shows the first five natural frequency modes and their effective mass participation factor (EMPF) in the X, Y, and Z directions.}
\label{fig:4}
\end{figure}

Figure \ref{fig:3} shows that any deflections on the optical plate are minimal with the max deflection being 11.62 \textmu m and the deflections at the mounting locations of critical optical components being lower. The figure shows deflection values at specific locations on the optical plate for OAP 1, the pupil wheel, OAP 2, the RIO tower, and the detector mount of 4.9, 3.6, 2.5, 4.7, and 1.3 \textmu m respectively. Each of these critical locations indicates that the deflections are within range for adjustability.

\subsection{Modal Analysis}\label{sec:2.2}

Similar to the static finite element analysis (FEA), for the modal or natural frequency analysis, the full CAD model was loaded into SolidWorks Simulation. The same point masses were used except for the RIO tower as a simplified model was included because its mode shape has a significant affect on the assembly's resulting natural frequencies. The bottom of the adapter frame was designated as the fixture point for the simulation. Gravity was once again the only external force in this analysis, and an average mesh density was used.

Figure \ref{fig:4} shows the results of the modal simulation with the lowest natural frequency mode on the left and a graph showing the effective mass participation factor (EMPF) versus mode on the right. The lowest mode is right on the edge of the Keck Observatory's keep out range of 8-80 Hz, but all other modes are significantly higher. This 80 Hz mode represents mass participation mostly at the RIO tower and along the x-direction. 

This 80 Hz result comes after already adjusting the design of the imager baffling to provide rigidity against vibration. The design of the baffling was originally described in (\citenum{Wiley2022}) where it prioritizes accessibility to underlying components. The main changes were to make the baffling thicker and to add additional fastener locations all around the optical plate including the filter wheel. The decreased accessibility is not significant while this has
\begin{wrapfigure}{r}{0.7\textwidth}
  \begin{center}
    \includegraphics[width=0.7\textwidth]{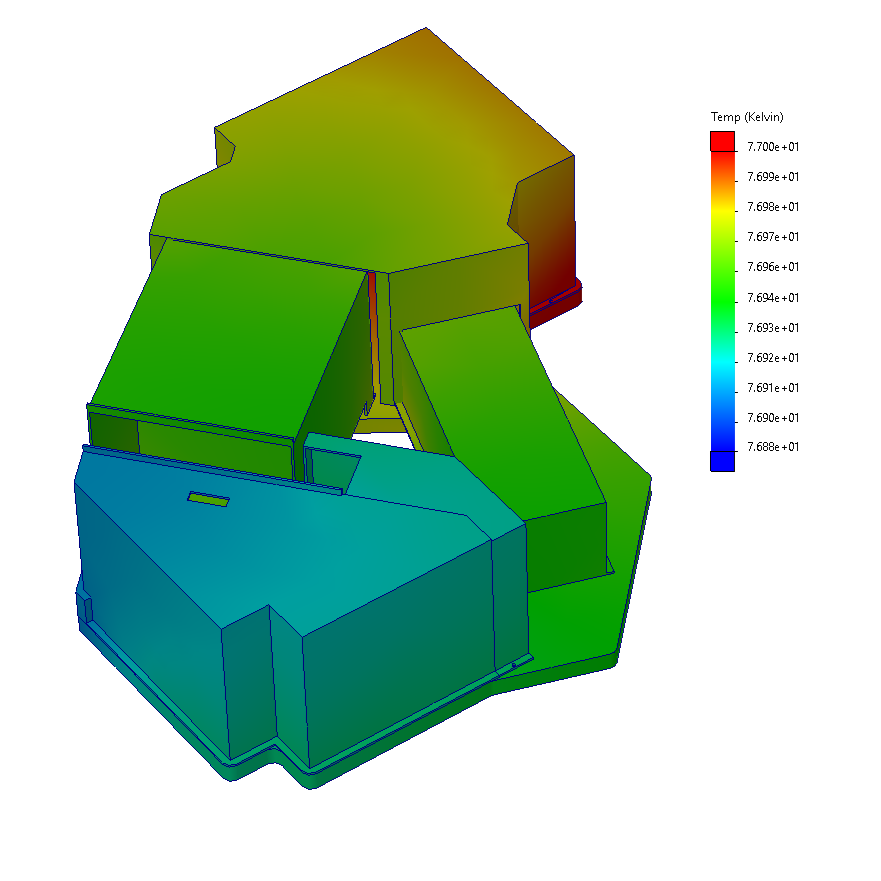}
  \end{center}
  \caption[Figure5]{SolidWorks Simulation result of the thermal analysis showing the temperature gradient across the imager optics plate. At 77 K, the temperature  difference across the optics plate is 0.12 K.}
\label{fig:5}
\end{wrapfigure}
shown to be an effective way of increasing the assembly's natural frequencies. To increase the lowest frequency mode further, the RIO tower will require additional stiffeners such as struts that attach to the baffling, or structural baffling of its own. 

\subsection{Thermal Analysis}\label{sec:2.3}

To ensure the proper operation of the imager, it is crucial to understand the temperature effects and gradient across the optical plate and its components. The imager model used for the thermal analysis is even more simplified than the static and modal analyses. This model removes all holes and changes all surfaces to be flat to simplify the radiation simulation. Each component is represented by a block with the identical thermal mass of that component. The imager is simulated in a box that represents the cold shield in the imager test chamber being cooled as described in (\citenum{Wiley2020}).

The thermal analysis set up uses the conductive and radiative load from the cold shield to the imager. Convection is negligible as the imager operates in vacuum. The temperature gradient across the cold shield is about 20 K between the lid and the base\cite{Wiley2020}. The simulation specifies the power output of the cold head such that the maximum temperature on the imager optical plate is 77 K.

The results in figure \ref{fig:5} show the temperature gradient across the imager optical plate being very small at around 0.12 K. With a linear thermal expansion coefficient of 7.8$\times$10$^{-6}$ K$^{-1}$ for 6061 T6 aluminum at 77 K\cite{NIST2006} and the largest distance on the optics plate of 1226mm, the maximum difference between two optics due to thermal expansion is only around 1 \textmu m. This static thermal gradient is well within the requirements for aligning the imager. 

The 0.12 K result is a small thermal gradient but can realistically be achieved due to the thickness of the optical plate providing high conductivity. The placement of copper straps between the optics plate and cold head is critical for maintaining this low thermal gradient. Despite any differences due to changes in temperature, the imager plate sits on kinematic mounts so it will rest in the same position between repeated cooldowns.

\section{Fabrication and Assembly Plan}\label{sec:3}

The Liger imager will be fabricated and assembled in the Optical and InfraRed Laboratory (OIRLab) at the University of California San Diego (UCSD). An in-lab characterization setup has been established for the Liger imager and similar future instruments. The imager will be cleanly assembled and warm aligned bench-top with a coordinate measuring machine (CMM) before being characterized inside a custom-built cryostat. Individual components will undergo their own cryogenic testing before being added to the full optical assembly. After characterization, the imager will be shipped to the University of California Los Angeles (UCLA) where it will be integrated with the Liger spectrograph and sciecnce cryostat.

The Liger imager characterization chamber was the first major component fabricated and is currently being characterized in the OIRLab. The chamber is floating on pneumatic isolators for vibration suppression and is attached to earthquake restraints for safety during seismic activity. The cold shield and other internal components are being cleaned before installation inside the vacuum chamber. The chamber is currently being pumped on to outgas as much as possible before the installation of additional components. Before delivery to the lab, the vacuum chamber was leak tested by the machine shop with no major leaks detected. The lowest vacuum achieved thus far is below $10^{-6}$ Torr and held when the pumps were turned off. This is important as the rough and turbo pumps will be turned off to decrease vibrations and the vacuum will be maintained by the cold head during normal operation.  Measurements of the reduction in vibrations, temperature gradient measurements, and residual gas analyses are yet to be performed.

Fabrication of individual imager components has commenced with drawings for the optics and detector mount being generated first, and requests for quotes being sent out. Drawings for the pupil wheel and filter wheel are currently being generated. As each component is fabricated and received in the lab, it will undergo ultrasonic and hand cleaning. Then, it will be passed in to the clean room for assembly. Some components such as those with motors will be individually characterized and cryogenically verified in the test chamber before integration with the full imager assembly.

This section describes the general in-lab setup and work-flow for assembling and characterizing the Liger imager and the fabrication status of its components. \S\ref{sec:3.1} details this work-flow and the area of the lab dedicated to this characterization setup. \S\ref{sec:3.2} provides further discussion on the current fabrication status of the Liger imager characterization chamber and optical assembly.

\subsection{Imager Assembly Work Flow}\label{sec:3.1}

\begin{figure}[t]
\begin{center}
  \includegraphics[width=0.8\linewidth,trim={0.0in 0.0in 0.0in 0.0in}]{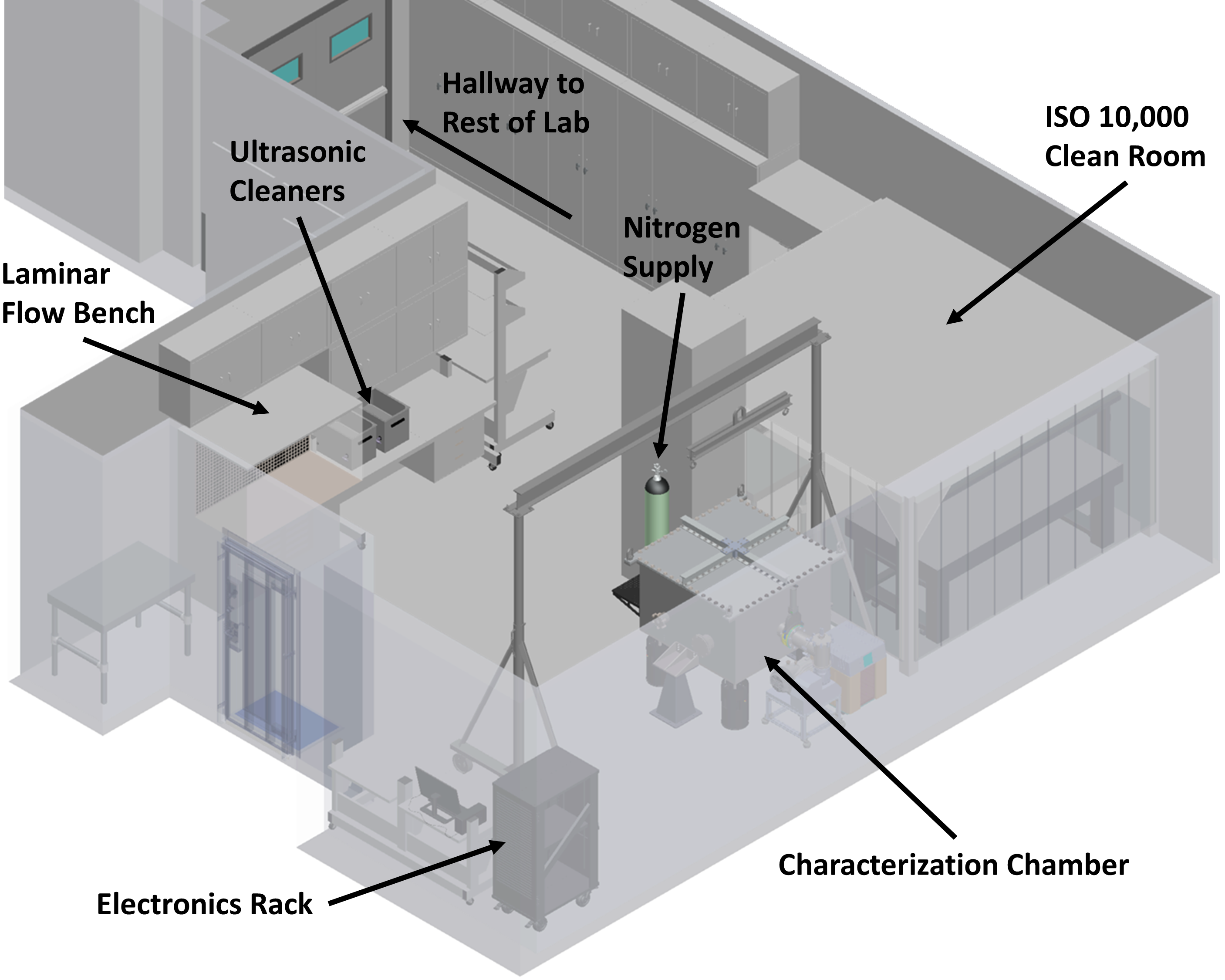}
  \end{center}
\caption[Figure6]{Annotated model of the room for optics characterization in the Optical Infrared Laboratory (OIRLab) at the University of California San Diego (UCSD) where the Liger imager will be assembled and characterized. This model includes the necessary work flow elements such as the laminar flow bench, cryostat, and clean room setups.}
\label{fig:6}
\end{figure}

One room of the OIRLab has been renovated to allow for the testing of cryogenic and temperature controlled Liger imager. This room for optics characterization has previously been used for the assembly and testing of instruments such as modules for the Panoramic Search for Extraterrestrial intelligence (PANOSETI)\cite{Maire2022} and the wavefront sensor replacement for the upgrade to the Gemini Planetary Imager (GPI 2.0)\cite{Perera2022}. 

The room for optics characterization consists of multiple optical tables, an ESD workstation, a setup for electronics, a 3D printer, and a clean room with a vibration suppressed optical table inside. For the assembly and characterization of the Liger imager and future such instruments, a cryogenic characterization chamber and supporting hardware was installed. Figure \ref{fig:6} shows a CAD model of the room where the characterization setup is shown with the chamber positioned under the 2 ton limit crane. The chamber is supported by an electronics rack and workstation in the corner of the room. The nitrogen supply brings the chamber up to atmospheric pressure while decreasing risk of contamination and condensation from room air\cite{Wiley2020}.

An overview of the work-flow as components are received is as follows. 1: Each individual component is cleaned by hand and/or in an ultrasonic bath with the proper cleaning agents (e.g. isopropyl alcohol, acetone). Components are further cleaned and inspected in the laminar flow bench. 2: Components are passed in to the ISO 10,000 clean room where larger parts and the imager as a whole will be assembled. 3: Critical imager components will be moved to the test cryostat for cryogenic validation before integration into the full imager optical assembly. 4: A vibrationally isolated optical table in the clean room will be the warm test bed for the Liger imager, where we will set up and test the optics. Alignment is performed utilizing a CMM for accurate positioning. Table \ref{tab:3} gives the tolerances that each optic must be aligned within during this process. 5: The imager optical assembly will be characterized in the test cryostat utilizing a Zygo interferometer and telescope simulator unit (TSU) for measuring and validating the alignment of the optical assembly when cold.


\begin{table}[t]
\centering
\caption{Mechanical tolerances of imager elements}
\label{tab:3}
\begin{tabular}{|c|c|c|c|} 
\hline
\textbf{Element} & \textbf{Decenter} (mm) & \textbf{Tilt} (degree/microns) & \textbf{Clocking} (degrees)\\\hline
OAP 1 & $\pm$ 0.100 & $\pm$ 0.075/42.5 & $\pm$ 0.500\\\hline
Fold Mirror & $\pm$ 0.500 & $\pm$ 0.050/19.0 & -\\\hline
OAP 2 & $\pm$ 0.100 & $\pm$ 0.050/48.0 & $\pm$ 0.500\\\hline
Detector & $\pm$ 0.100 & $\pm$ 0.250 & $\pm$ 0.200 \\\hline
Pick-Offs & $\pm$ 0.050 & $\pm$ 0.200 & $\pm$ 0.026\\\hline
\end{tabular}
\end{table}

\subsection{Fabrication}\label{sec:3.2}

\begin{figure}[b]
\begin{center}
  \includegraphics[width=0.8\linewidth,trim={0.0in 0.0in 0.0in 0.0in}]{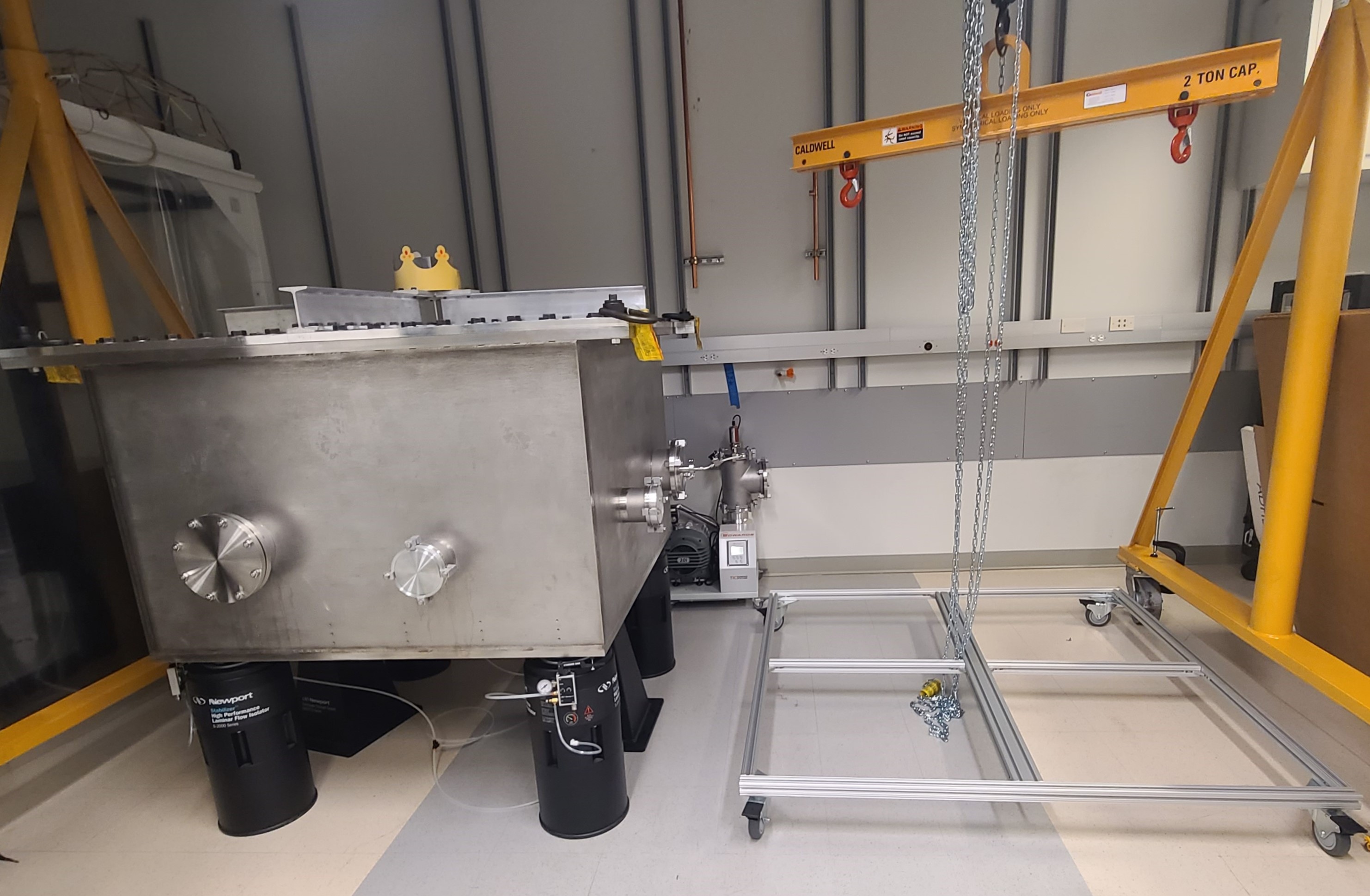}
  \end{center}
\caption[Figure7]{Picture taken of the vacuum chamber set up in the OIRLab. The vacuum chamber is mounted on pneumatic isolators and placed under the crane. The cart for removing the lid and cold shield is placed next to the chamber under the spreader bar.}
\label{fig:7}
\end{figure}

The Liger imager finished its final design phase in late 2023 and has commenced its initial fabrication phase\cite{Brown_2024}. Of components being fabricated for the imager, the test cryostat was the first to be finished and delivered to the lab. This was planned to allow for ample time to generate comprehensive procedures for using the characterization setup and practice before testing critical components. 

The vacuum chamber was fabricated by UCSD's Machining and Additive Prototyping Services (MAPS). The chamber is constructed of welded together steel walls and flanges with an aluminum lid\cite{Wiley2020}. The inside is polished to a low surface roughness to decrease vacuum outgassing and thermal emissivity. The chamber underwent leak testing and verification by MAPS before delivery to the OIRLab.

After delivery, the 2600 lbs chamber was installed on the pneumatic isolators using the crane for lifting. A picture showing this setup is given in figure \ref{fig:7} where the vacuum chamber is resting on the pneumatic isolators under the crane next to the clean room. The space next to the chamber is clear, allowing for room to remove the lid and other components.

Smaller elements of the test cryostat are currently being fabricated such the cold head attachment which is being machined at the Qualcomm Institute's Prototyping Lab. The CaFl entrance window, which has a 60 mm clear aperture and is mounted on the NW-100 flange, will be sent for fabrication soon. The electrical feedthroughs for the test chamber are mounted on the other NW-100 and 160 flanges and will be machined soon as well. 

The next components to be fabricated for the imager optical assembly are the detector mount and the optics. Fabrication drawings have been generated for the flat mirror, both OAPs, and the detector mount, and have been sent for quotes. Once these components are fabricated and received, we will align just the optics on the bench in the clean room for warm testing of any components before assembly on the optics plate.

\section{Summary \& Future Actions}\label{sec:4}

Liger is a state-of-the-art facility-class instrument that will provide advancements for a wide range of science cases. It takes design heritage from the previous and next generation instruments OSIRIS and IRIS, generating a powerful new hybrid instrument for Keck Observatory. The improvements from Liger are necessary for Keck to fully take advantage of the enhancements to Keck I AO provided by KAPA. The imager subsystem of Liger is critical to the whole instrument as it subsequently feeds both IFS subsystems. In this paper, the structural integrity of the imager was analyzed, and the fabrication and characterization plan was reviewed.

Simulations mostly confirm that the imager meets requirements with the modal analysis showing the lowest resonant frequency mode at 80 Hz, and most mass participation being near the RIO tower. The static analysis confirms the structural integrity of the assembly and shows that any deflections from the expected position of each major component is minimal at less than 12$\mu$m. The thermal analysis shows a small temperature gradient of 0.12 K across the optical plate. This leads to a very small deflection due to thermal expansion across the imager optical plate.

The imager has finished its final design phase and commenced fabrication with the first major component delivered being the test cryostat. One room of the OIRLab has been renovated for the general characterization of optics. The test chamber was installed in this room along with other necessary elements for the assembly and characterization of the imager. Drawings of the imager components are currently being generated and will be sent out for fabrication upon completion. As parts arrive in the lab, they will be cleaned, assembled, and characterized.

In the upcoming stages of establishing the lab's characterization setup and fabricating the imager, several critical steps are planned. These include conducting vibration measurements, performing a residual gases analysis, and assessing temperature gradients in the cryostat, and measuring the air particle count in the clean room. The design of the RIO is being finalized along with the baffling and support structure around it.

The fabrication and characterization of the imager are pivotal to the overall success of the Liger instrument. The detailed assembly and characterization plan is essential to guarantee the precision, reliability, and optimal performance of the Liger imager. This comprehensive approach ensures that every component is thoroughly tested and validated, increasing the reliability and longevity of the instrument. The superior performance of the imager will sequentially provide the IFS with high quality photons, advancing the observational capabilities of Liger and W.M. Keck Observatory\cite{Kassis2022}.

\acknowledgments 
The design and development of the Liger instrument have been made possible by the Heising-Simons Foundation (Grant 2018-1085), the Gordon and Betty Moore Foundation (Grant 11169), and the National Science Foundation (NSF) Advanced Technologies and Instrumentation (ATI Grant 2308190). 

\bibliography{liger} 

\begin{thebibliography}{10}

\bibitem{Wizinowich2022}
{Wizinowich}, P., {Lu}, J.~R., {Cetre}, S., {Chin}, J., {Correia}, C., {Delorme}, J.~R., {Gers}, L., {Lilley}, S., {Lyke}, J., {Marin}, E., {Ragland}, S., {Richards}, P., {Surendran}, A., {Wetherell}, E., {Chen}, C.~F., {Chu}, D., {Do}, T., {Fassnacht}, C., {Freeman}, M., {Gautam}, A., {Ghez}, A., {Hunter}, L., {Jones}, T., {Liu}, M.~C., {Mawet}, D., {Max}, C., {Morris}, M., {Phillips}, M., {Ruffio}, J.~B., {Rundquist}, N.~E., {Sabhlok}, S., {Terry}, S., {Treu}, T., and {Wright}, S., ``{Keck All sky Precision Adaptive optics program overview},'' in [{\em Adaptive Optics Systems VIII}{\nolinebreak\hspace{0.1em}]},  {Schreiber}, L., {Schmidt}, D., and {Vernet}, E., eds., {\em Society of Photo-Optical Instrumentation Engineers (SPIE) Conference Series} {\bf 12185},  121850Q (Aug. 2022).

\bibitem{Wright2019}
{Wright}, S., {Larkin}, J.~E., {Jones}, T., {Kupke}, R., {Fitzgerald}, M., {Kassis}, M., {Cosens}, M., {Chisholm}, E., {Do}, T., {Fassnacht}, C., {Fisher}, D., {Ghez}, A., {Johnson}, C., {Keane}, J., {Kirby}, E., {Kress}, E., {Konopacky}, Q., {Lu}, J.~R., {Maire}, J., {O'Meara}, J., {Reddy}, N., {Sanders}, R., {Sandstrom}, K., {Shapley}, A., {Sohn}, J.-M., {Surya}, A., {Treu}, T., {Weber}, R., {Wiley}, J., {Wizinowich}, P., {Wong}, M., and {Zonca}, A., ``{Liger: Next Generation Imager and Spectrograph for Keck Observatory Adaptive Optics},'' in [{\em Bulletin of the American Astronomical Society}{\nolinebreak\hspace{0.1em}]},   {\bf 51},  201 (Sept. 2019).

\bibitem{Larkin2006}
{Larkin}, J., {Barczys}, M., {Krabbe}, A., {Adkins}, S., {Aliado}, T., {Amico}, P., {Brims}, G., {Campbell}, R., {Canfield}, J., {Gasaway}, T., {Honey}, A., {Iserlohe}, C., {Johnson}, C., {Kress}, E., {LaFreniere}, D., {Lyke}, J., {Magnone}, K., {Magnone}, N., {McElwain}, M., {Moon}, J., {Quirrenbach}, A., {Skulason}, G., {Song}, I., {Spencer}, M., {Weiss}, J., and {Wright}, S., ``{OSIRIS: a diffraction limited integral field spectrograph for Keck},'' in [{\em Society of Photo-Optical Instrumentation Engineers (SPIE) Conference Series}{\nolinebreak\hspace{0.1em}]},  {McLean}, I.~S. and {Iye}, M., eds., {\em Society of Photo-Optical Instrumentation Engineers (SPIE) Conference Series} {\bf 6269},  62691A (June 2006).

\bibitem{Ghez2008}
{Ghez}, A.~M., {Salim}, S., {Weinberg}, N.~N., {Lu}, J.~R., {Do}, T., {Dunn}, J.~K., {Matthews}, K., {Morris}, M.~R., {Yelda}, S., {Becklin}, E.~E., {Kremenek}, T., {Milosavljevic}, M., and {Naiman}, J., ``{Measuring Distance and Properties of the Milky Way's Central Supermassive Black Hole with Stellar Orbits},'' {\em The Astrophysical Journal}~{\bf 689},  1044--1062 (Dec. 2008).

\bibitem{Larkin2020}
{Larkin}, J.~E., {Wright}, S.~A., {Chisholm}, E.~M., {Andersen}, D., {Dekany}, R.~G., {Dunn}, J.~S., {Hayano}, Y., {Kupke}, R., {Smith}, R., {Suzuki}, R., {Weber}, R.~W., and {Zhang}, K., ``{The Infrared Imaging Spectrograph (IRIS) for TMT: instrument overview},'' in [{\em Ground-based and Airborne Instrumentation for Astronomy VIII}{\nolinebreak\hspace{0.1em}]},  {Evans}, C.~J., {Bryant}, J.~J., and {Motohara}, K., eds., {\em Society of Photo-Optical Instrumentation Engineers (SPIE) Conference Series} {\bf 11447},  114471Y (Dec. 2020).

\bibitem{Wiley2022}
{Wiley}, J., {Brown}, A., {Kupke}, R., {Cosens}, M., {Wright}, S.~A., {Fitzgerald}, M., {Johnson}, C., {Jones}, T., {Kassis}, M., {Kress}, E., {Larkin}, J.~E., {Magnone}, K., {McGurk}, R., {Rundquist}, N., {Wang}, E., and {Yeh}, S., ``{Liger at Keck Observatory: design of imager optical assembly and spectrograph re-imaging optics},'' in [{\em Ground-based and Airborne Instrumentation for Astronomy IX}{\nolinebreak\hspace{0.1em}]},  {Evans}, C.~J., {Bryant}, J.~J., and {Motohara}, K., eds., {\em Society of Photo-Optical Instrumentation Engineers (SPIE) Conference Series} {\bf 12184},  1218464 (Aug. 2022).

\bibitem{Cosens2022}
{Cosens}, M., {Wright}, S.~A., {Brown}, A., {Fitzgerald}, M., {Johnson}, C., {Jones}, T., {Kassis}, M., {Kress}, E., {Kupke}, R., {Larkin}, J.~E., {Magnone}, K., {McGurk}, R., {Rundquist}, N.-E., {Sohn}, J.~M., {Wang}, E., {Wiley}, J., and {Yeh}, S., ``{Liger at Keck Observatory: image detector and IFS pick-off mirror assembly},'' in [{\em Ground-based and Airborne Instrumentation for Astronomy IX}{\nolinebreak\hspace{0.1em}]},  {Evans}, C.~J., {Bryant}, J.~J., and {Motohara}, K., eds., {\em Society of Photo-Optical Instrumentation Engineers (SPIE) Conference Series} {\bf 12184},  121845Y (Aug. 2022).

\bibitem{Wiley2020}
{Wiley}, J., {Mathur}, K., {Brown}, A., {Wright}, S., {Cosens}, M., {Maire}, J., {Fitzgerald}, Michael amd~{Jones}, T., {Kassis}, M., {Kress}, E., {Kupke}, R., {Larkin}, J.~E., {Lyke}, J., {Wang}, E., and {Yey}, S., ``{Liger for Next-Generation Keck Adaptive Optics: Cryogenic Chamber for the Imaging Camera},'' in [{\em Ground-based and Airborne Instrumentation for Astronomy VI}{\nolinebreak\hspace{0.1em}]},  {Shields}, J., ed., {\em Society of Photo-Optical Instrumentation Engineers (SPIE) Conference Series} {\bf 11447},  11447--311 (Dec. 2020).

\bibitem{Kupke_2024}
{Kupke}, R., {Wright}, S.~A., {Brown}, A., {Cosens}, M., {Fitzgerald}, M., {Johnson}, C., {Jones}, T., {Kassis}, M., {Kress}, E., {Larkin}, J.~E., {Magnone}, K., {Maire}, J., {McGurk}, R., {Surya}, A., {Wang}, E., {Wiley}, J., and {Yeh}, S., ``{Liger at W.M. Keck Observatory: optical design and alignment strategy for the Liger integral field spectrograph},'' Presented at SPIE Astronomical Telescopes + Instrumentation 2024. Forthcoming publication (July 2024).

\bibitem{Cosens2020}
{Cosens}, M., {Wright}, S.~A., {Arriaga}, P., {Brown}, A., {Fitzgerald}, M., {Jones}, T., {Kassis}, M., {Kress}, E., {Kupke}, R., {Larkin}, J.~E., {Lyke}, J., {Wang}, E., {Wiley}, J., and {Yeh}, S., ``{Liger for next-generation Keck AO: filter wheel and pupil design},'' in [{\em Ground-based and Airborne Instrumentation for Astronomy VIII}{\nolinebreak\hspace{0.1em}]},  {Evans}, C.~J., {Bryant}, J.~J., and {Motohara}, K., eds., {\em Society of Photo-Optical Instrumentation Engineers (SPIE) Conference Series} {\bf 11447},  114474X (Dec. 2020).

\bibitem{Wright_2024}
{Wright}, S. A. e.~a., ``{Liger at W. M. Keck Observatory: overall design and fabrication status},'' Presented at SPIE Astronomical Telescopes + Instrumentation 2024. Forthcoming publication (July 2024).

\bibitem{Brown_2024}
{Brown}, A., {Wright}, S., {Wiley}, J., {Cosens}, M., {Kupke}, R., {Kassis}, M.~F., {McGurk}, R., {Yeh}, S., {Larkin}, J.~E., {Fitzgeral}, M.~P., {Wang}, E., {Kress}, E., {Magnone}, K., {Johnson}, C., and {Jones}, T., ``{Liger at W.M. Keck Observatory: assembly, integration, and testing},'' (July 2024).
\newblock Presented at SPIE Astronomical Telescopes + Instrumentation 2024. Forthcoming publication.

\bibitem{Sabhlok2024}
{Sabhlok}, S., {Wright}, S.~A., {Lu}, J.~R., {Terry}, S., {Wizinowich}, P.~L., {Neichel}, B., and {Kuznetsov}, A., ``{On-axis point spread function reconstruction performance validation for Keck NIRC2},'' Presented at SPIE Astronomical Telescopes + Instrumentation 2024. Forthcoming publication (July 2024).

\bibitem{Vayner2016}
{Vayner}, A., {Wright}, S.~A., {Do}, T., {Larkin}, J.~E., {Armus}, L., and {Gallagher}, S.~C., ``{Providing Stringent Star Formation Rate Limits of z {\ensuremath{\sim}} 2 QSO Host Galaxies at High Angular Resolution},'' {\em The Astrophysical Journal}~{\bf 821},  64 (Apr. 2016).

\bibitem{NIST2006}
Division, N.~C., ``Cryogenic material properties database,'' tech. rep., National Institute of Standards and Technology (2006).
\newblock Update 2006.

\bibitem{Maire2022}
{Maire}, J., {Wright}, S.~A., {Holder}, J., {Anderson}, D., {Benbow}, W., {Brown}, A., {Cosens}, M., {Foote}, G., {Hanlon}, W.~F., {Hervet}, O., {Horowitz}, P., {Howard}, A.~W., {Lee}, R., {Liu}, W., {Raffanti}, R., {Rault-Wang}, N., {Stone}, R. P.~S., {Werthimer}, D., {Wiley}, J., and {Williams}, D.~A., ``{Panoramic SETI: program update and high-energy astrophysics applications},'' in [{\em Ground-based and Airborne Instrumentation for Astronomy IX}{\nolinebreak\hspace{0.1em}]},  {Evans}, C.~J., {Bryant}, J.~J., and {Motohara}, K., eds., {\em Society of Photo-Optical Instrumentation Engineers (SPIE) Conference Series} {\bf 12184},  121848B (Aug. 2022).

\bibitem{Perera2022}
{Perera}, S., {Maire}, J., {Do {\'O}}, C.~R., {Nguyen}, J.~S., {Levinstein}, D.~M., {Konopacky}, Q.~M., {Chilcote}, J., {Fitzsimmons}, J., {Hamper}, R., {Kerley}, D., {Macintosh}, B., {Marois}, C., {Rantakyr{\"o}}, F., {Savransky}, D., {Veran}, J.-P., {Agapito}, G., {Ammons}, S.~M., {Bonaglia}, M., {Boucher}, M.-A., {Dunn}, J., {Esposito}, S., {Filion}, G., {Landry}, J.~T., {Lardiere}, O., {Li}, D., {Dillon}, D., {Madurowicz}, A., {Peng}, D., {Poyneer}, L., and {Spalding}, E., ``{GPI 2.0: pyramid wavefront sensor status},'' in [{\em Adaptive Optics Systems VIII}{\nolinebreak\hspace{0.1em}]},  {Schreiber}, L., {Schmidt}, D., and {Vernet}, E., eds., {\em Society of Photo-Optical Instrumentation Engineers (SPIE) Conference Series} {\bf 12185},  121854C (Aug. 2022).

\bibitem{Kassis2022}
{Kassis}, M.~F., {Allen}, S., {Alvarez}, C., {Banyal}, R., {Bertz}, R., {Beichman}, C., {Brown}, A., {Brown}, M., {Cabak}, G., {Bundy}, K., {Cetre}, S., {Chin}, J., {Chun}, M., {Deich}, W., {Dekany}, R., {Delorme}, J., {Devenot}, M., {Doppmann}, G., {Fitzgerald}, M.~P., {Fucik}, J.~R., {Hill}, G., {Hinz}, P., {Holden}, B.~P., {Howard}, A., {Gao}, M., {Gibson}, S., {Gomez}, P., {Gottschalk}, C., {Gillingham}, P.~R., {Jones}, T., {Jovanovic}, N., {Kirby}, E., {Konopacky}, Q., {Krishnan}, S., {Kupke}, R., {Larkin}, J.~E., {Leifer}, S.~D., {Lewis}, H.~A., {Lilley}, S., {Lu}, J., {Lyke}, J.~E., {MacDonald}, N., {Marin}, E., {Matuszewski}, M., {Mawet}, D., {McGurk}, R., {Millar-Blanchaer}, M.~A., {Nash}, R.~B., {Nance}, C., {Neill}, J.~D., {O'Meara}, J.~M., {Peretz}, E., {Poppett}, C.~L., {Mather}, J.~C., {Radovan}, M.~V., {Roberts}, M.~K., {Ragland}, S., {Rider}, K., {Rockosi}, C.~M., {Sandfor}, D., {Shen}, B., {Steidel}, C.~C., {Simha}, S., {Skemer}, A., {Stelter}, D., {Surendran}, A., {Thorne}, J., {McCarney},
  B., {Lanclos}, K., {Baker}, A., {Rubenzahl}, R., {Roy}, A., {Halverson}, S., {Edelstein}, J., {Martin}, C., {Savage}, M., {Sandford}, D., {Sallum}, S., {Walawender}, J., {Wizinowich}, P., {Westfall}, K.~B., {Vahala}, K.~J., {Wright}, S., {Wold}, T., and {Yeh}, S., ``{Innovations and advances in instrumentation at the W. M. Keck Observatory, vol. II},'' in [{\em Ground-based and Airborne Instrumentation for Astronomy IX}{\nolinebreak\hspace{0.1em}]},  {Evans}, C.~J., {Bryant}, J.~J., and {Motohara}, K., eds., {\em Society of Photo-Optical Instrumentation Engineers (SPIE) Conference Series} {\bf 12184},  1218405 (Aug. 2022).

\end{thebibliography}
\bibliographystyle{spiebib} 

\end{document}